\documentclass[aps,prl,twocolumn,superscriptaddress]{revtex4}

\usepackage{graphicx}
\usepackage{dcolumn}
\usepackage{bm}
\usepackage{color}
\usepackage{epstopdf}
\usepackage{natbib}
\usepackage{amssymb}
\usepackage{textcomp}
\usepackage{mathtools}
\newcommand{\ch}[1]{\textcolor{black}{#1}} 

\begin{document}

\title{Sensing the spin of an individual Ce adatom}

\author{Markus Ternes}
\affiliation{RWTH Aachen University, Institute of Physics, 
D-52074 Aachen, Germany}
\affiliation{Peter-Gr\"{u}nberg-Institute, Forschungszentrum J\"{u}lich, 
D-52425 J\"{u}lich, Germany}

\author{Christopher P. Lutz}
\affiliation{IBM Almaden Research Center, San Jose, California 95120,
USA}

\author{Andreas J. Heinrich}
\affiliation{Center for Quantum Nanoscience, Institute for Basic Science (IBS), 
Seoul 03760, Republic of Korea}
\affiliation{Physics Department, Ewha Womans University,
Seoul, Republic of Korea}

\author{Wolf-Dieter Schneider}
\affiliation{Ecole Polytechnique F\'ed\'erale de Lausanne (EPFL), Institut de 
Physique,
CH-1015 Lausanne, Switzerland}
\affiliation{Fritz-Haber-Institute of the Max-Planck-Society, Faradayweg 4-6, 
D-14195 Berlin, Germany}

\date{\today}

\begin{abstract}

The magnetic moment of rare earth elements originates from electrons in the 
partially filled 4$f$ orbitals. Accessing this moment electrically by scanning 
tunneling spectroscopy is hampered by 
shielding 
of outer-lying orbitals. Here we show that we 
can detect the magnetic moment of an individual Ce atom adsorbed on a Cu$_2$N 
ultrathin film on Cu(100) by using a sensor tip that has its apex 
functionalized 
with a Kondo screened spin system. We calibrate the sensor tip by 
deliberately coupling it to a well characterized Fe atom.  Subsequently, we use 
the splitting of the tip's Kondo resonance when approaching a spectroscopically 
dark Ce atom to sense its magnetic moment.

\end{abstract}

\maketitle

Recently, the magnetism of surface-supported rare earth elements has come newly
into focus, because individual atoms with $4f$ electrons on ultrathin 
insulators have been found to show long relaxation times \cite{Donati16}, 
making 
them interesting candidates for atomic scale memory and possible qubit 
realization. Due to their large orbital angular momentum, which results in less 
extended orbitals than the $3d$ orbitals of transition metals like Fe or Co, 
$f$ orbitals do not usually take part in chemical bonds and only hybridize 
weakly. This isolation is both a blessing and a curse as this promotes magnetic 
stability, but at the cost of easy detection and manipulation by electrical 
means. 	

Accordingly, scanning tunneling microscopy (STM) and spectroscopy measurements 
on Ho and Gd adatoms on Pt(111) revealed only low 
or no detectable interaction cross section between the tunneling electrons and 
the localized spin \cite{Schuh12, Balashov14, Steinbrecher16}. Nevertheless, 
the $4f$ moment of Ho atoms adsorbed on a thin insulating film of MgO on 
Ag(100) was detected as a change in the electron spin resonance frequency of a 
single Fe adatom \cite{Natterer17} and changed the spectrum of a Co atom in 
HoCo dimers \cite{Singha18}.

Furthermore, in compounds and thin layers of Ce, a lanthanide which hosts only 
one 4$f$ electron, Kondo screening has been observed \cite{Patthey85, 
Patthey87, 
Patthey90,Laubschat90,Ehm07}. In Kondo systems, the magnetic moments get 
compensated by itinerant electrons 
leading to a highly correlated singlet state and a resonance 
at the Fermi energy below a characteristic Kondo 
temperature $T_K$ \cite{Kondo64, Kondo68, Hewson97, Ternes09}. Kondo 
features have been found on surface-supported double-decker molecules 
containing Dy \cite{Warner16}, however, measurements on Ce atoms on Ag(111) 
showed ambiguous results. While 
first results hinted at Kondo screening \cite{Li98a}, subsequent investigations 
revealed that single Ce adatoms diffuse on the Ag(111) surface even at 
$4.7$~K \cite{Silly04,Silly04a,Negulyaev09, Ternes10} suggesting that the 
earlier measurement was taken on an immobile Ce cluster and that hydrogenated 
4$f$ atoms can show low-energy vibrational excitations mimicking a spin signal 
\cite{Pivetta07}. However, small Ce clusters 
showed a clear Kondo resonance \cite{Ternes09}.

\begin{figure*}[]
\includegraphics[width=1.6\columnwidth]{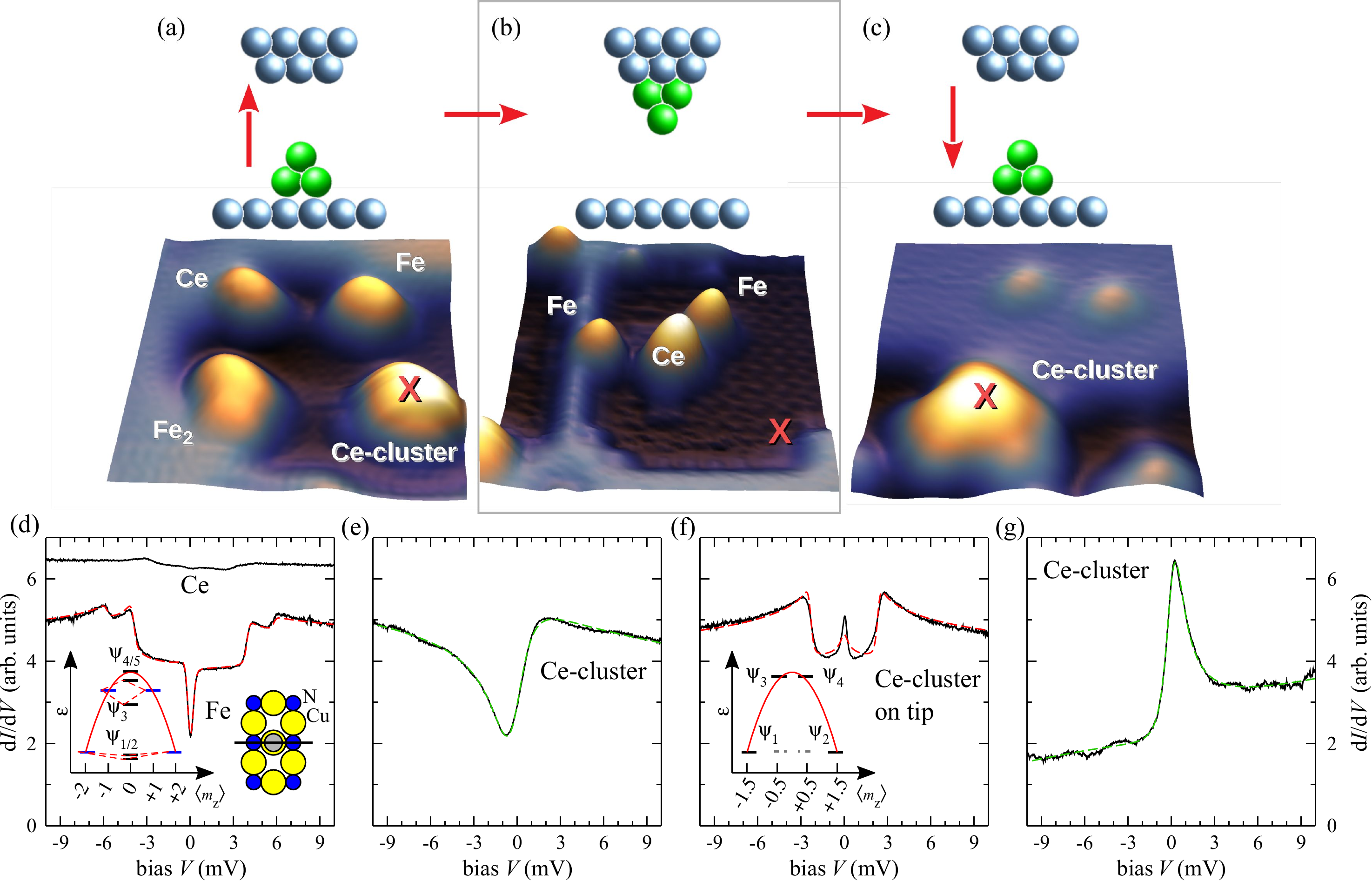}
\caption{(color online) \textbf{(a-c)} Schemes and 
STM images ($5\times 5$\,nm$^2$, $V=10$\,mV, $I=1$\,nA) illustrating the 
formation of the functionalized tip by picking up 
(a$\rightarrow$b) and dropping down (b$\rightarrow$c) a small Ce cluster. 
\textbf{(d-g)} Spectra (black lines) and fits using the 
scattering model (dashed red lines) or a Fano line-shape (dashed green lines).
(d) Spectrum of a Fe atom with spin $S=2$ and axial anisotropy 
along the nitrogen rows of the Cu$_2$N surface (right inset) with 
$D=-1.6$\,meV, 
$E=0.3$\,meV (left inset: state diagram), $J\rho_0=-0.09$ and $U=0.35$ 
(bottom), and a Ce atom with flat spectrum (top). (e) Ce cluster showing Kondo 
resonance having $\Gamma=1.3\pm0.1$\,meV, $q\approx0.45$, 
and $T_K\approx 15$\,K (cross in (a)). (f) Same Ce cluster after transfer to 
the 
tip-apex with $S=3/2$ 
(inset: state diagram)
measured against the bare surface (cross in (b)) and (g) after transfer back 
to the surface showing again a Kondo resonance having 
$\Gamma=0.80\pm0.02$\,meV, $q\approx3$, and $T_K\approx 9.3$\,K.}
\label{fig-1}
\end{figure*}

Therefore, we re-addressed the question of the magnetic moment of single Ce 
adatoms by co-depositing individual Fe and Ce atoms onto a monolayer of 
Cu$_{2}$N on  Cu(100) \cite{Leibsle94} and performing experiments with a 
home-built STM at $T_{\rm exp}=0.7$~K. As shown in Fig.~\ref{fig-1}a, single Ce 
and Fe 
atoms as well as small clusters formed during deposition at $\approx 
10$~K are imaged as stable protrusions. The single atoms 
adsorb on the Cu-sites of the Cu$_{2}$N \cite{SOM}. 
Measuring the differential conductance 
$dI/dV$ 
of Fe atoms revealed 
characteristic features originating from spin-flip 
excitations 
\cite{Heinrich04, Hirjibehedin07}. To interpret these spectra 
we use a model Hamiltonian for the magnetic atom,
\begin{equation}\label{Hamiltonian}
\hat{H}_{\rm imp}= 
D\hat{S}_{z}^2+E(\hat{S}_{x}^2-\hat{S}_{y}^2)-g\mu_B\vec{B}\hat{
\boldsymbol S},
\end{equation} 
where the spin system is described by the generalized spin operator 
$\hat{\boldsymbol S}=(\hat{S}_x,\hat{S}_y,\hat{S}_z)$, $D$ and 
$E$ are the axial and transverse magnetic anisotropy parameters. The Zeeman 
energy is accounted for with $g\approx2$ as the gyromagnetic factor, $\mu_B$ 
the Bohr magneton, and $\vec{B}$ the applied magnetic field.

The step-like increase in $dI/dV$ is due to excitations from the ground 
to energetically higher eigenstates $\left| \Psi \right\rangle$ of 
Equ.~\ref{Hamiltonian} via Kondo-like interactions between the localized 
magnetic moment and the tunneling electron. The tunneling electron has 
states $\left|\phi\right\rangle$ and spin matrices 
$\hat{\boldsymbol\sigma}=(\hat{\sigma}_x,\hat{\sigma}_y,\hat{\sigma}_z)$,
leading to the transition matrix elements between initial ($i$) and final ($f$) 
states:
\begin{equation}\label{Interaction}
M_{i\rightarrow f}=\sum_{i',f'}\left\langle \phi_{f'},\Psi_{f} \right| 
\left( \frac12
\hat{\boldsymbol\sigma}\cdot\hat{\boldsymbol 
S}+U\hat{\sigma}_0\cdot\hat{1} 
\right) \left| \phi_{i'},\Psi_{i} \right\rangle.
\end{equation} 
Here, 
$\left| \phi_{i'},\Psi_{i} \right\rangle=\left| \phi_{i'}\right\rangle\left| 
\Psi_{i} \right\rangle$ are product states, $\hat{\sigma}_0$ and $\hat{1}$ 
are identity matrices in their corresponding Hilbert sub-spaces, and $U$ is a 
Coulomb scattering parameter which accounts not only for a background 
$dI/dV$, but also leads to interference induced bias asymmetries in 
the spectra, when higher scattering orders are considered \cite{Ternes15}. 

We find an excellent fit to the Fe data using the previously found 
effective spin $S=2$, easy-axis anisotropy ($D<0$) which favors the high 
magnetic moment along the N-rows of the Cu$_2$N \cite{Hirjibehedin07}, 
and a transport model which includes scattering processes up to 3$^{\rm rd}$ 
order in the matrices \cite{Ternes15, Ternes17} (Fig.~\ref{fig-1}d). For 
3$^{\rm rd}$ order processes, we additionally take the dimensionless coupling 
$J\rho_0$ into account, where $J$ is the interaction strength between the many 
electrons of the substrate and the atom's magnetic moment and 
$\rho_0$ is the density of states at Fermi energy. 

Interestingly, while the $dI/dV$ of Fe atoms show strong steps that 
indicate spin excitations, single Ce atoms do not. Their spectra are 
essentially flat and featureless (Fig.~\ref{fig-1}d). This changes when 
measurements are taken on a small cluster of Ce atoms \cite{Ternes09}. 
Fig.~\ref{fig-1}e shows that we detect an asymmetric feature centered near 
$V=0$ which can be well described by a Fano line shape \cite{Fano61, 
Madhavan01}:
\begin{equation}\label{Fano-equ}
dI/dV\propto\frac{(q+\epsilon')^2}{
1+\epsilon'^2 }+(1+\alpha\epsilon')\rho_0.
\end{equation} 
In this equation, $\epsilon'=(eV-\epsilon_K)/\Gamma$ is the normalized 
energy where $\epsilon_K\approx0$ is the position and $\Gamma$ the 
half-width of the resonance, $q$ is the Fano parameter, and 
$(1+\alpha\epsilon')\rho_0$ accounts for a background $dI/dV$. We 
find $\Gamma\approx1.3$\,meV which is due to Kondo screening of 
the cluster's magnetic moment with $T_K=\Gamma/k_B\approx 15$\,K. The 
$q\approx 0.45$ indicates a relatively strong direct tunneling channel between 
the tip and the substrate \cite{Ujsaghy00}. Because $T\ll T_K$ the cluster is 
in 
the strong screening regime \cite{Zonda18}. 
\begin{figure*}[t]
\includegraphics[width=1.5\columnwidth]{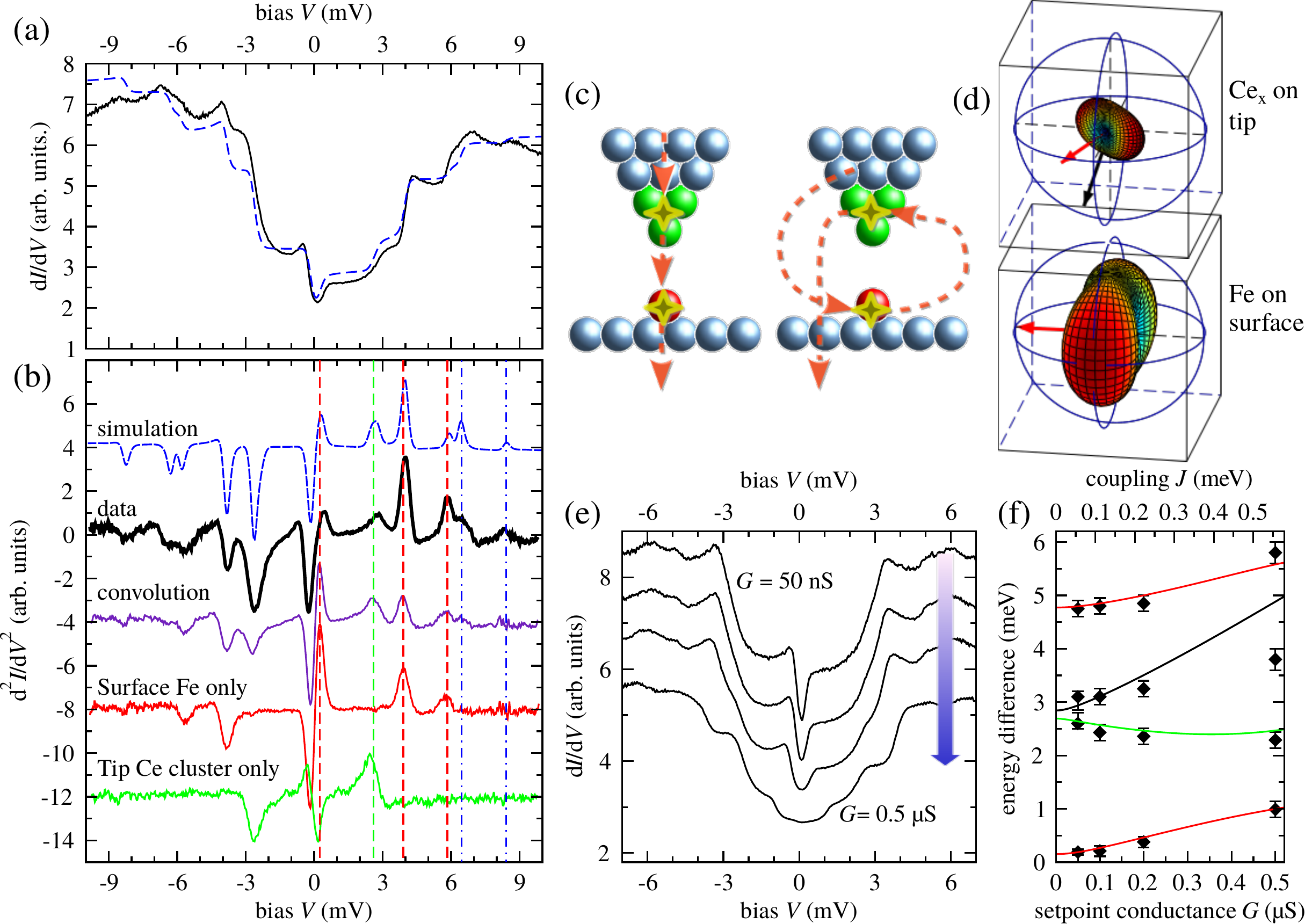}
\caption{(color online) \textbf{(a, b)} Spectra measured with the Ce 
cluster tip against the Fe adatom of Fig.~\ref{fig-1}d 
(black lines) and simulated curve (dashed blue line). For comparison, spectra 
of the Ce cluster tip, the Fe adatom, and the convolution of both are 
shown (vertically shifted colored lines in (b)). Vertical dashed lines mark 
transition energies. \textbf{(c)} Scheme of the interaction processes for an 
electron tunneling from tip to sample. 
\textbf{(d)} Visualization of the magnetic anisotropy of the Ce cluster 
and the Fe/Cu$_2$N surface spin. Red arrows mark the easy axes, while the back 
arrow mark the intermediate axis of the tip. \textbf{(e)} Set-point dependent 
spectra (top to bottom: $G=50, 100, 200, 500$\,nS) over a different Fe atom 
having slightly increased anisotropy parameters ($D=-1.85$\,meV, $E=0.3$\,meV). 
\textbf{(f)} Extracted step positions (symbols) and transition energies of the 
model calculations using $J^{\rm ts} \propto G$ (lines).}
\label{fig-3}
\end{figure*}

Surprisingly, when we transfer this Ce cluster to the STM tip, by applying 
$+1$~V bias pulses at close to point contact, the spectrum changes 
drastically, now revealing a narrow Kondo resonance, and a spin excitation at 
about $\pm2.5$~meV (Fig.~\ref{fig-1}f). This spectrum is typical for an 
effective $S=3/2$ system in which the zero-bias Kondo peak originates from 
scattering between the two degenerate ground states that have weights in 
the $m_z=\pm1/2$ projections, and the steps from transitions to 
states that are energetically higher by $\Delta\epsilon= 2\sqrt{D^2+3E^2}$ 
(inset Fig.~\ref{fig-1}f). Similar to Fe, this spectrum can be fit with 
the scattering model even though the central peak is only partly represented, 
since Kondo screening includes multiple higher-order scattering. 
This spectrum is very similar to that of a single Co atom on Cu$_{2}$N 
\cite{Otte08a, Oberg13, Bergmann15}, but here the magnetic moment originates 
from the Ce cluster at the apex of the metallic tip, enabling us to exploit 
it as a mobile sensor as shown below.

Transferring the cluster back to the surface as a control experiment by moving 
the tip to near point contact and withdrawing while applying 
$V=-1$~V leads to a similar spectrum as before pick-up 
(Fig.~\ref{fig-1}g). Both spectra 
show a Kondo signal, but the transfer of the cluster to a new location on 
Cu$_2$N has changed the Kondo signal from dip to a peak. However, $\Gamma$ is 
changed only modestly, with $T_K$ changing by about $40\%$, which might be due 
to 
different adsorption geometries \cite{Gao07}. 
In contrast to the relatively weak coupling of the cluster when attached to the 
tip apex, the apparently stronger coupling to bulk electrons on the surface 
quenches the magnetic anisotropy \cite{Oberg13, Jacobson15} and moves the spin 
system into the strong-coupling Kondo regime. 

To characterize the magnetic properties of the Ce cluster at the tip apex, we 
probe Fe atoms using the Ce-functionalized tip, which results in a complex 
spectrum (Fig.~\ref{fig-3}a, b). Interestingly, the second derivative 
$d^2I/dV^2$ closely resembles the convolution of the individual spectra of tip 
and sample, similar to recent observations of coupled molecular spins in a 
junction \cite{Ormaza17}. However, wFhile the convolution yields the 
correct transition energies, the peak intensities differ significantly between 
data and convolution (Fig.~\ref{fig-3}b). 

This difference has two origins: First, transport which excites both spin 
systems is complex because the scattering events can occur in different 
sequences, which must be summed coherently. An electron tunneling from tip (t) 
to sample (s) can scatter first with the spin on the tip and then with the spin 
on the surface, and it can also first interact with the surface spin before 
interacting with the tip spin (Fig.~\ref{fig-3}c). Destructive quantum 
interference between these different scattering channels cancel all scattering 
processes, except those that obey 
$[\hat{\sigma}_i^{\rm t},\hat{\sigma}_j^{\rm s}]_{+}\neq0$. For spins on tip 
and surface, 
this results in a different form for the interaction matrix than 
Equ.~\ref{Interaction}, giving \cite{SOM}:
\begin{align}\label{Interaction-two}
M_{i\rightarrow f}=\!\!
\sum_{\substack{j=x,y,z \\i',f'}}\!\!\left\langle 
\phi_{f'},\Psi_{f} \right|&\bigg[
U^{\rm s}\hat{\sigma}_j(\hat{S}_j^{\rm t}\otimes\hat{1}^{\rm s})
+U^{\rm t}\hat{\sigma}_j(\hat{1}^{\rm t}\otimes\hat{S}_j^{\rm s})
\nonumber\\
&+\frac12\hat{\sigma}_0(\hat{S}_j^{\rm t}\otimes\hat{S}_j^{\rm s})\bigg]
\left| \phi_{i'},\Psi_{i} \right\rangle.
\end{align}
Here, the first two terms account for scattering in which spin-spin interaction
occurs at only one of the two localized moments, while on the other only a 
Coulomb-like interaction take place. The last term accounts for 
spin-spin scattering on both the tip and sample moments.

Second, while the anisotropy axis of the Fe adatom is well known, the 
anisotropy axis of the Ce cluster on the tip is unknown and might point in an 
arbitrary direction. To determine this direction, we use a model that employs 
Equ.~\ref{Interaction-two} up to 3$^{\rm rd}$ order and a Hamiltonian 
$\hat{H}=\hat{H}_{\rm Ce_x}+\hat{H}_{\rm Fe}$ in which the Ce cluster and the 
Fe adatom are both described by Equ.~\ref{Hamiltonian} using 
\ch{their individually found anisotropy and coupling parameters \cite{SOM}.} 
This model allows us to determine 
the relative 
alignment of the two spins. As illustrated in Fig.~\ref{fig-3}d we find a 
\ch{$\approx70$\textdegree}\ angle between the two magnetic easy axes. 
Furthermore, 
the Ce cluster's magnetic intermediate axis, that is the $y$-direction in
Equ.~\ref{Hamiltonian}, is tilted by only $\approx 18$\textdegree\ from the 
surface normal. Note, that the simplicity of the model limits its accuracy 
\ch{\cite{SOM}}.

To determine the magnetic coupling between the tip and surface spins, we 
proceed 
by changing the set-point conductance $G=I/V$ and consequently the tip-sample 
distance $z$. We observe a shift in intensity and energy of the excitations 
(Fig.~\ref{fig-3}e). In particular the energy of the first excitation changes 
from $0.2$\,meV to $1.0$\,meV, when $G$ is varied by one order of magnitude, 
i.\,e.~when $z$ is 
reduced by $\approx 1$\AA. This shift can be well described by assuming an 
antiferromagnetic Heisenberg-like exchange interaction 
$J^{\rm ts}\hat{\boldsymbol 
S}^{\rm t} \cdot \hat{\boldsymbol S}^{\rm s}$ between both spins 
(Fig.~\ref{fig-3}f). 
The linear dependence, $J^{\rm ts}=(1.1\pm0.2)\mbox{meV}/\mu\mbox{S}\times G$, 
implies an exponential dependence on $z$ and points to an orbital overlap as 
origin of the interaction \cite{Muenks17, Verlhac19, Yang19}.

\begin{figure}[]
\includegraphics[width=1\columnwidth]{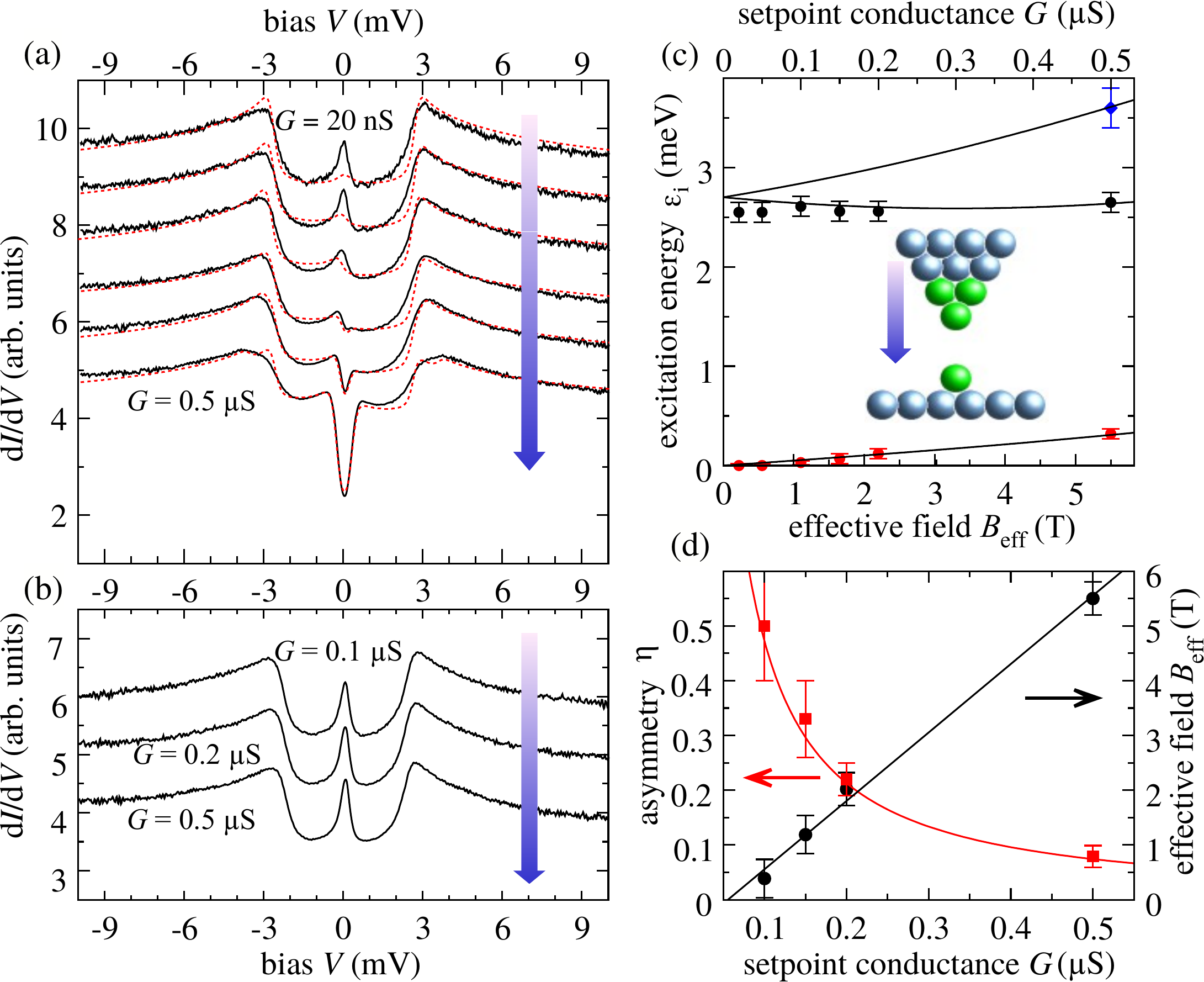}
\caption{(color online) \textbf{(a)} Spectra (black lines) measured with the 
functionalized Ce cluster tip positioned above a 
``dark'' Ce adatom with increasing conductance 
(top to bottom: $G=20,50,100,150,200,500$\,nS) and simulations (dashed red 
lines) using a perturbation model ($S=3/2$, $D=-1.3$\,meV, 
$E=0.18$\,meV, $J\rho_0=-0.17$) and a $G$ dependent effective $B$ field.
\textbf{(b)} Same tip measured against the bare Cu$_2$N surface.
\textbf{(c)} Extracted step positions (symbols) and calculated transition 
energies (lines) for different $G$ and $B$ values, 
respectively. The inset shows schematically the experimental set-up. 
\textbf{(d)} From the data in (a) extracted asymmetry $\eta$ of the central 
peak and $B$ field. The lines are guides for the eyes.}
\label{fig-2}
\end{figure}

Having carefully characterized the magnetic tip, we use it to probe 
spectroscopically ``dark'' Ce adatoms. Fig.~\ref{fig-2}a shows typical spectra, 
which for small $G$ (tip far from the surface) are almost identical to the ones 
measured against the bare surface (Fig.~\ref{fig-2}a, b) except for a a slight 
reduction of the zero bias Kondo peak height. Prominent differences emerge at 
increased $G$. We observe that the central peak becomes asymmetric and for 
$G\geq 0.15\,\mu$S splits 
while spectra taken at similar $G$ values on the bare 
surface do not change significantly (Fig.~\ref{fig-2}b).

This result clearly reveals the presence of a magnetic moment in the Ce 
adatom that influences the spectrum of the probing tip. We can model the 
observed behavior by using Equ.~\ref{Hamiltonian} and assuming that the 
spectroscopically dark Ce adatom acts as an effective, set-point dependent, 
magnetic field oriented in a direction approximately out-of-surface having a 
strength of $B_y\approx (11\pm1)$\,T/$\mu{\rm S}\times G$ 
(Fig.~\ref{fig-2}c, d). 

While part of the spectral asymmetry can be attributed to 
Fano-like interference (Equ.~\ref{Fano-equ}), we interpret the asymmetric 
intensity of the split central feature to originate primarily from an
imbalance 
$\eta=\left(\rho_0(\uparrow\right)-\rho_0(\downarrow)) / (\rho_0(\uparrow 
)+\rho_0 ( \downarrow))$ between majority $(\uparrow)$ and minority 
$(\downarrow)$ sample states \cite{Loth10b, Bergmann15}. 
Such polarization can be induced by the spin moment of the Ce adatom 
\cite{Muenks17, 
Verlhac19}. The decrease with $G$ suggests an antiferromagnetic 
interaction as the origin of the exchange field $B$ and 
the formation of a combined non-magnetic singlet state of both spins. This 
indicates a half-integer moment of the Ce adatom and assuming $S=1/2$, the 
Ising-like coupling leads to a strength of $J_y^{\rm ts}=(2.5\pm 
0.3)$\,meV/$\mu$S$\times G$. We found that using isotropic Heisenberg or 
dipole-dipole interactions instead of Ising interactions lead to much less 
adequate fits. We also have indications of a reduced tip-adatom interaction 
strength for Ce adsorbed close to the bare metal \cite{SOM}.

To summarize, our results point to an effective spin $S=1/2$ of single Ce 
adatoms on the Cu$_2$N surface. The very localized $4f$ orbital is inaccessible 
to ordinary spectroscopy because the interaction cross section 
with the tunneling electron is too small. This is also the reason for the 
absence of Kondo screening at our accessible temperatures. Nevertheless, we 
sensed the magnetic $4f$ moment by a ``detector'' spin at the apex of a 
functionalized STM tip. The particular coupling mechanism between the two spins 
is not yet fully known, but the observed exponential dependence of the 
Ising-like interaction points to some $4f-5d$ orbital mixing 
\cite{Gunnarsson83}. 

As a detector we used a Ce cluster which showed a very narrow Kondo resonance 
at $E_F$. 
Such a tip is ideally suited 
to spin detection, in particular, when $T_K$ is of order $T_{\rm 
exp}$. Then, the tip shows a clear Kondo resonance at $E_F$ and its sensitivity 
to exchange interactions and spin polarization is only limited by the 
unavoidable thermal broadening of the spectra.
In contrast, tips with high $T_K > T_{\rm exp}$ or very low $T_K \ll T_{\rm 
exp}$ are contra-indicative. While in the former the exchange 
interaction has first to overcome the Kondo energy $k_BT_K$ before spectral 
changes can be observed \cite{Bork11}, the latter don't show a pronounced 
resonance. Note, however, magnetic molecules at 
the tip apex can be exploited in a similar manner as sensor and may be prepared 
more reproducibly \cite{Verlhac19, Czap19}.

Our results open a new route for studying $4f$ elements on well-defined 
surfaces. For example, the method outlined here could be used to measure the 
interactions in artificially created or self-assembled \cite{Silly04, Ternes10} 
atomic or molecular 1D and 2D $4f$ nanostructures.

\begin{acknowledgments}
MT acknowledges support by the Heisenberg Program
(Grant No.\ TE 833/2-1) of the German Science Foundation and AJH from the 
Institute for Basic Science under grant IBS-R027-D1. We thank H.-J.\ Freund 
and A.\ Singha for stimulating discussions.

\end{acknowledgments}


\end{document}